\documentstyle[amssymb,preprint,aps]{revtex}

\begin{document}
\author{A.S. Alexandrov}
\address{Department of Physics, Loughborough University, Loughborough LE11\\
3TU, United Kingdom}
\title{Bloch waves of small high-Tc bipolarons}
\maketitle

\begin{abstract}
Over the last decade several competing models of high-temperature
superconductivity were proposed, most of them with short-range interactions.
We review a more realistic model with strong on-site repulsive correlations,
the Coulomb and strong finite-range electron-phonon interactions. Bipolarons
in the model exist in the itinerant Bloch states at temperatures below about
half of the characteristic phonon frequency. Depending on the ratio of the
inter-site Coulomb repulsion and the polaron level shift the ground state of
the model is a polaronic Fermi (or Luttinger) liquid, bipolaronic high-$%
T_{c} $ superconductor, or charge-segregated insulator for the strong,
intermediate, and weak Coulomb repulsion, respectively. Two particular
lattices are analysed in detail: a chain with the finite range
electron-phonon interaction and a zig-zag ladder. Charge carriers in the
ladder are superlight mobile intersite bipolarons. They propagate coherently
without emission or absorption of phonons with about the same mass as single
polarons. The model describes key features of the cuprates, in particular
their Tc values, different isotope effects, normal state pseudogaps, and
spectral functions measured in tunnelling and photoemission.

74.20.-z,74.65.+n,74.60.Mj
\end{abstract}

\section{Introduction}

For although high-temperature superconductivity has not been yet targeted as
`{\it the shame and despair of theoretical physics}', - a label attributed
to superconductivity during the first half-century after its discovery - the
parlous state of current theoretical constructions has led to a current
consensus that there is no consensus on the theory of high-$T_{c}$
superconductivity\cite{aleedw}. Our view is that the extension of the BCS
theory towards the strong interaction between electrons and ion vibrations
describes naturally the phenomenon, and the high temperature
superconductivity exists in the crossover region of the electron-lattice
interaction strength from the BCS-like to bipolaronic superconductivity \cite
{ale0}. Quite remarkably Bednorz and M\"{u}ller noted in their original
publication \cite{bed} and subsequently in their Nobel Price lecture \cite
{bed2}, that in their ground-breaking search for high-T$_{c}$
superconductivity, they were stimulated and guided by the polaron model.
Their expectation \cite{bed2} was that if `{\it an electron and a
surrounding lattice distortion with a high effective mass can travel through
the lattice as a whole, and a strong electron-lattice coupling exists an
insulator could be turned into a high temperature superconductor}'. Indeed
there is now overwhelming experimental \cite{guo,shen,ega,mul2} and
theoretical \cite{alemot,dev,allen,gor,bis2,dev2} evidence for an
exceptionally strong electron-phonon (e-ph) interaction in high temperature
superconductors. Thus the theory of HTSC must include both e-ph and
electron-electron Coulomb interactions as it was suggested some time ago\cite
{alemot}. Also one has to take into account that all oxides are highly
polarizable ionic lattices. A low density of mobile carriers is unable to
screen effectively the direct Coulomb electron-ion and electron-electron
interactions. The layered structure of the cuprates reduces screening even
further. Since the mobile carriers are confined to the copper-oxygen planes
their interaction with c-axis polarized optical phonons cannot be screened,
and it is particularly strong. The parameter-free estimate of the polaron
binding energy, $E_{p}$ due to the long-range Fr\"{o}hlich e-ph interaction
puts it at about 0.5 eV or larger in the cuprates \cite{alebra}.

However bipolaronic states are much heavier than band electrons since they
are `dressed' by the same lattice deformation, which bounds two polarons in
a pair \cite{alemot}. As a result, the superconducting critical temperature, 
$T_{c}$, being proportional to the inverse mass of a bipolaron, might be
significantly reduced rather than enhanced compared with the weak-coupling
BCS $T_{c}.$ Because of this prejudice some objections have been raised with
respect to the bipolaron theory of high-temperature superconductivity. \ In
the present article we review a few recent models of small bipolarons with
the long-range electron-phonon (e-ph) interaction capturing the essential
physics of superconducting cuprates \cite{alebook}. These studies prove that
small bipolarons are perfectly mobile Bloch states in the cuprates which can
explain their high $T_{c}$.

The electron-phonon coupling constant $\lambda $ is about the ratio of the
electron-phonon interaction energy $E_{p}$ to the half bandwidth $D$ in a
rigid lattice. We expect that when the coupling is strong, $\lambda >1$, all
electrons in the Bloch band are ``dressed'' by phonons because their kinetic
energy ($<D$) is small compared with the potential energy due to a local
lattice deformation caused by an electron. If phonon frequencies are very
low, the local lattice deformation traps the electron. This {\it %
self-trapping }phenomenon was predicted by Landau \cite{lan}. It has been
studied in greater detail by Pekar and Fr\"{o}hlich, and later on the most
advanced path-integral theory of polarons was developed by Feynman and
Devreese and his school in the effective mass approximation, which leads to
the so-called {\it large polaron} (for more detail see Ref. \cite{dev}). The
large polaron propagates through the lattice like a free electron but with
the enhanced effective mass. In the strong-coupling regime, $\lambda >1,$
the finite bandwidth becomes important, so that the effective mass
approximation cannot be applied. The electron is called a {\it small polaron 
}in this regime. The self-trapping is never ``complete'', that is any
polaron can tunnel through the lattice coherently. Only in the extreme {\it %
adiabatic} limit, when the phonon frequencies tend to zero, the
self-trapping is complete, and the polaron motion is no longer
translationally continuous. The main features of the small polaron were
understood by Tjablikov \cite{tja}, Yamashita and Kurosava \cite{yam},
Sewell \cite{sew}, Holstein and his school \cite{hol}, Lang and Firsov \cite
{fir}, and others and described in several review papers and textbooks \cite
{app,fir2,alemot,dev,bry,mah,alemot2}. The exponential reduction of the
bandwidth at large values of $\lambda $ is one of those features. The small
polaron bandwidth decreases with increasing temperature up to a crossover
region from the coherent small polaron tunneling to a thermally activated
hopping. The crossover from the polaron Bloch states to the incoherent
hopping takes place at temperatures $T\approx \omega _{0}/2$ or higher,
where $\omega _{0}$ is the characteristic phonon frequency. Here we show
that {\it small \ bipolarons} are also in the itinerant Bloch states in a
wide temperature region. Moreover, they propagate with about the same mass
as a single polaron in particular lattices such as perovskites.

\section{ Small polaron band}

The canonical approach to a small polaron problem is based on the
displacement (Lang-Firsov) transformation of the electron-phonon Hamiltonian 
\cite{fir} in the site (${\bf m)}$ representation for electrons allowing for
the summation of all diagrams including the vertex corrections, 
\begin{eqnarray}
H &=&\sum_{i,j}T({\bf m-n})\delta _{ss^{\prime }}c_{i}^{\dagger }c_{j}+ \\
\sum_{{\bf q},i}\omega _{{\bf q}}\hat{n}_{i}\left[ u_{i}({\bf q})d_{{\bf q}%
}+H.c.\right] &+&{\frac{1}{{2}}}\sum_{i\neq j}V_{c}({\bf m-n})\hat{n}_{i}%
\hat{n}_{j}+\sum_{{\bf q}}\omega _{{\bf q}}(d_{{\bf q}}^{\dagger }d_{{\bf q}%
}+1/2).  \nonumber
\end{eqnarray}
Here $T({\bf m})$ is the bare hopping integral, 
\[
u_{i}({\bf q})={\frac{1}{\sqrt{2N}}}\gamma ({\bf q})e^{i{\bf q\cdot m}} 
\]
is the matrix element of the electron-phonon interaction, $i=({\bf m},s)$, $%
j=({\bf n},s^{\prime })$, $s=\uparrow \downarrow $, $\hat{n}%
_{i}=c_{i}^{\dagger }c_{i}$, $c_{i},d_{{\bf q}}$ are the electron (hole) and
phonon operators, respectively, and $N$ is the number of sites ($\hslash
=c=k_{B}=1).$

Following Lang and Firsov \cite{fir} one can apply the canonical
transformation $e^{S}$ to diagonalise the Hamiltonian. The diagonalisation
is exact if $T({\bf m})=0$ (or $\lambda =\infty $): 
\begin{equation}
\tilde{H}=e^{S}He^{-S},
\end{equation}
where 
\begin{equation}
S=-\sum_{{\bf q},i}\hat{n}_{i}\left[ u_{i}({\bf q})d_{{\bf q}}-H.c.\right] 
\end{equation}
The electron operator transforms as 
\begin{eqnarray}
\tilde{c}_{i} &=&c_{i}\hat{X}_{i}, \\
\hat{X}_{i} &=&\exp \left[ \sum_{{\bf q}}u_{i}({\bf q})d_{{\bf q}}-H.c.%
\right] ,  \nonumber
\end{eqnarray}
and the phonon one as 
\begin{equation}
\tilde{d}_{{\bf q}}=d_{{\bf q}}-\sum_{i}\hat{n}_{i}u_{i}^{\ast }({\bf q}).
\end{equation}
It follows from Eq.(5) that the Lang-Firsov canonical transformation shifts
ions to new equilibrium positions. In a more general sense it changes the
boson vacuum. As a result, 
\begin{eqnarray}
\tilde{H} &=&\sum_{i,j}\hat{\sigma}_{ij}c_{i}^{\dagger }c_{j}-E_{p}\sum_{i}%
\hat{n}_{i}+  \nonumber \\
&&\sum_{{\bf q}}\omega _{{\bf q}}(d_{{\bf q}}^{\dagger }d_{{\bf q}}+1/2)+{%
\frac{1}{{2}}}\sum_{i\neq j}v_{ij}\hat{n}_{i}\hat{n}_{j},
\end{eqnarray}
where 
\begin{equation}
\hat{\sigma}_{ij}=T({\bf m-n})\delta _{ss^{\prime }}\exp \left( \sum_{{\bf q}%
}[u_{j}({\bf q})-u_{i}({\bf q})]d_{{\bf q}}-H.c.\right) 
\end{equation}
is the renormalised hopping integral depending on the phonon variables, and 
\begin{equation}
v_{ij}=V_{c}({\bf m-n})-{\frac{1}{{N}}}\sum_{{\bf q}}|\gamma ({\bf q}%
)|^{2}\omega _{{\bf q}}\cos [{\bf q\cdot (m-n})]
\end{equation}
is the the interaction of polarons owing to the Coulomb repulsion $V_{c}(%
{\bf m-n})$ and to the local lattice deformation (the second term).

In an extreme strong coupling limit $\lambda \rightarrow \infty $ one can
neglect the hopping term of the transformed Hamiltonian. The rest has
analytically determined eigenstates and eigenvalues. The eigenstates $|%
\tilde{N}\rangle =|n_{i},n_{{\bf q}}\rangle $ are classified with the
polaron $n_{{\bf m},s}$ and phonon $n_{{\bf q}}$ occupation numbers and the
energy levels are 
\begin{equation}
E=-E_{p}\sum_{i}n_{i}+{\frac{1}{{2}}}\sum_{i\neq j}v_{ij}n_{i}n_{j}+\sum_{%
{\bf q}}\omega _{{\bf q}}(n_{{\bf q}}+1/2)
\end{equation}
where $n_{i}=0,1$ and $n_{{\bf q}}=0,1,2,3,....\infty $.

Hence, the Hamiltonian Eq.(1) in zero order with respect to the hopping
describes localised polarons and independent phonons which are vibrations of
ions relative to new equilibrium positions depending on the polaron
occupation numbers. The phonon frequencies remain unchanged in this limit.
The middle of the electronic band falls by the polaronic level shift $E_{p}$
as a result of a potential well created by the lattice deformation, 
\begin{equation}
E_{p}={\frac{1}{{2N}}}\sum_{{\bf q}}|\gamma ({\bf q})|^{2}\omega _{{\bf q}}.
\end{equation}

With the finite hopping term polarons tunnel in a narrow band owing to the
degeneracy of the zero order Hamiltonian with respect to the site position
of a single polaron in a regular lattice. To see it one can apply the
perturbation theory using $1/\lambda $ as a small parameter with 
\begin{equation}
\lambda \equiv \frac{E_{p}}{zT(a)},
\end{equation}
where $z$ is the coordination lattice number and $T(a)$ is the
nearest-neighbour hopping integral, so that $D\approx zT(a)$. The proper
(Bloch) set of $N$ degenerate zero order eigenstates of the lowest energy
level ($-E_{p}$) of the unperturbed Hamiltonian is 
\begin{equation}
|{\bf k},0\rangle ={\frac{1}{\sqrt{N}}}\sum_{{\bf m}}c_{{\bf m}}^{\dagger
}\exp (i{\bf k\cdot m})|0\rangle ,
\end{equation}
$|0\rangle $ is the vacuum. By applying the textbook perturbation theory one
readily calculates the lowest energy levels of the polaron in a crystal. Up
to the second order in the hopping integral the result is 
\begin{eqnarray}
E({\bf k}) &=&-E_{p}+\epsilon _{{\bf k}} \\
- &&\sum_{{\bf k^{\prime }},n_{{\bf q}}}{\frac{|\langle {\bf k},0|\sum_{i,j}%
\hat{\sigma}_{ij}c_{i}^{\dagger }c_{j}|{\bf k^{\prime }},n_{{\bf q}}\rangle
|^{2}}{{\sum_{{\bf q}}\omega _{{\bf q}}n_{{\bf q}}}}}  \nonumber
\end{eqnarray}
where $|{\bf k^{\prime }},n_{{\bf q}}\rangle $ are exited states of the
unperturbed Hamiltonian with one electron and at least one real phonon. The
second term in Eq.(13), which is linear with respect to the bare hopping,
determines the small polaron band dispersion as 
\begin{equation}
\epsilon _{{\bf k}}=\sum_{{\bf m}}t({\bf m})\,e^{-g^{2}({\bf m})}\exp (-i%
{\bf k\cdot m}),
\end{equation}
with the band-narrowing factor (at zero temperature) 
\begin{equation}
g^{2}({\bf m})={\frac{1}{{2N}}}\sum_{{\bf q}}|\gamma ({\bf q})|^{2}[1-\cos (%
{\bf q\cdot m})]
\end{equation}
The third term in Eq. (13), quadratic in $T(a)$, yields a negative ${\bf k}$-%
{\em independent} correction to the polaron level shift of the order of $%
1/\lambda ^{2}$. 

\section{ Small bipolaron band}

The attractive energy of two small polarons, $2D(\lambda -\mu _{c})$ is
generally larger than the polaron bandwidth in the strong-coupling regime, 
\begin{equation}
\lambda -\mu _{c}\gg Z^{\prime }
\end{equation}
because the bandwidth narrowing factor $Z^{\prime }$ is small 
\[
Z^{\prime }={\frac{\sum_{{\bf m}}T({\bf m})e^{-g^{2}({\bf m})}\exp (-i{\bf %
k\cdot m})}{{\sum_{{\bf m}}T({\bf m})\exp (-i{\bf k\cdot m})}}\ll 1,}
\]
where $\mu _{c}$ is the Coulomb pseudopotential. When this condition is
fulfilled, small bipolarons are not overlapped. Hence the polaronic Fermi
liquid transforms into a Bose liquid of double-charged carriers. Here we
encounter a novel electronic state of matter, a charged Bose liquid \cite
{ale0,aleedw}, qualitatively different from the normal Fermi-liquid and from
the BCS superfluid.

\subsection{Onsite bipolaron band}

The small parameter, $Z^{\prime }/(\lambda -\mu _{c})\ll 1,$ allows for a
consistent treatment of bipolaronic systems \cite{ale0,aleran}. Under this
condition the hopping term in the transformed Hamiltonian $\tilde{H}$ is a
small perturbation of the ground state of immobile bipolarons and free
phonons, 
\begin{equation}
\tilde{H}=H_{0}+H_{pert},
\end{equation}
where 
\begin{equation}
H_{0}={\frac{1}{{2}}}\sum_{i,j}v_{ij}c_{i}^{\dagger }c_{j}^{\dagger
}c_{j}c_{i}+\sum_{{\bf q}}\omega _{{\bf q}}[d_{{\bf q}}^{\dagger }d_{{\bf q}%
}+1/2]
\end{equation}
and 
\begin{equation}
H_{pert}=\sum_{i,j}\hat{\sigma}_{ij}c_{i}^{\dagger }c_{j}
\end{equation}
Let us first discuss dynamics of {\it onsite }\ bipolarons, which are the
ground state of the system with the Holstein-type non-dispersive e-ph
interaction. The onsite bipolaron is formed if 
\begin{equation}
2E_{p}>U,
\end{equation}
where $U$ is the onsite Coulomb correlation energy (the so-called Hubbard $U$%
). The intersite polaron-polaron interaction Eq.(8) is purely Coulomb
repulsion because the phonon mediated attraction between two polarons on
different sites is zero in the Holstein model. Two or more onsite bipolarons
as well as three or more polarons cannot occupy the same site because of the
Pauli exclusion principle. Hence, bipolarons repel single polarons and each
other. Their binding energy, $\Delta =2E_{p}-U,$ is normally larger than the
polaron half-bandwidth, $\Delta \gg w=Z^{\prime }D,$ so that there are no
unbound polarons in the ground state. $H_{pert}$, Eq.(19), destroys
bipolarons in the first order. Hence it has no diagonal matrix elements.
Then the bipolaron dynamics, including superconductivity, is described by
the use of a new canonical transformation $\exp (S_{2})$ \cite{aleran},
which eliminates the first order of $H_{pert}$, 
\begin{equation}
(S_{2})_{fp}=\sum_{i,j}{\frac{\langle f|\hat{\sigma}_{ij}c_{i}^{\dagger
}c_{j}|p\rangle }{{E_{f}-E_{p}}}}.
\end{equation}
Here $E_{f,p}$ and $|f\rangle ,|p\rangle $ are the energy levels and the
eigenstates of $H_{0}$. Neglecting the terms of higher orders than $%
(w/\Delta )^{2}$ we obtain 
\begin{equation}
(H_{b})_{ff^{\prime }}\equiv \left( e^{S_{2}}\tilde{H}e^{-S_{2}}\right)
_{ff^{\prime }},
\end{equation}
\begin{eqnarray*}
(H_{b})_{ff^{\prime }} &\approx &(H_{0})_{ff^{\prime }}-{\frac{1}{{2}}}%
\sum_{\nu }\sum_{i\neq i^{\prime },j\neq j^{\prime }}\langle f|\hat{\sigma}%
_{ii^{\prime }}c_{i}^{\dagger }c_{i^{\prime }}|p\rangle \langle p|\hat{\sigma%
}_{jj^{\prime }}c_{j}^{\dagger }c_{j^{\prime }}|f^{\prime }\rangle \times  \\
&&\left( {\frac{1}{{E_{p}-E_{f^{\prime }}}}}+{\frac{1}{{E_{p}-E_{f}}}}%
\right) .
\end{eqnarray*}
$S_{2}$ couples a localised onsite bipolaron and a state of two unbound
polarons on different sites. The expression (22) determines the matrix
elements of the transformed{\it \ bipolaronic} Hamiltonian $H_{b}$ in the
subspace $|f\rangle ,|f^{\prime }\rangle $ with no single (unbound)
polarons. On the other hand the intermediate {\it bra} $\langle p|$ and {\it %
ket} $|p\rangle $ in Eq.(22) refer to configurations involving two unpaired
polarons and any number of phonons. Hence we have 
\begin{equation}
E_{p}-E_{f}=\Delta +\sum_{{\bf q,\nu }}\omega _{{\bf q\nu }}\left( n_{{\bf %
q\nu }}^{p}-n_{{\bf q\nu }}^{f}\right) ,
\end{equation}
where $n_{{\bf q\nu }}^{f,p}$ are phonon occupation numbers $%
(0,1,2,3...\infty )$. This equation is an explicit definition of the
bipolaron binding energy $\Delta $ which takes into account the residual
intersite repulsion between bipolarons and between two unpaired polarons.
The lowest eigenstates of $H_{b}$ are in the subspace, which has only doubly
occupied $c_{{\bf m}\uparrow }^{\dagger }c_{{\bf m}\downarrow }^{\dagger
}|0\rangle $ or empty $|0\rangle $ sites. Onsite bipolaron tunnelling is a
two-step transition. It takes place via a single polaron tunneling to a
neighbouring site. The subsequent tunnelling of its ``partner'' to the same
site restores the initial energy state of the system. There are no $real$
phonons emitted or absorbed because the bipolaron band is narrow. Hence we
can average $H_{b}$ with respect to phonons. Replacing the energy
denominators in the second term in Eq.(22) by the integrals with respect to
time, 
\[
\frac{1}{{E_{p}-E_{f}}}=i\int_{0}^{\infty }dte^{i({E_{f}-Ep}+i\delta )t},
\]
we obtain 
\begin{eqnarray}
H_{b} &=&H_{0}-i\sum_{{\bf m\neq m}^{\prime },s}\sum_{{\bf n\neq n}^{\prime
},s^{\prime }}T({\bf m-m}^{\prime })T({\bf n-n}^{\prime })\times  \\
&&c_{{\bf m}s}^{\dagger }c_{{\bf m}^{\prime }s}c_{{\bf n}s^{\prime
}}^{\dagger }c_{{\bf n}^{\prime }s^{\prime }}\int_{0}^{\infty }dte^{-i\Delta
t}\Phi _{{\bf mm}^{\prime }}^{{\bf nn}^{\prime }}(t),  \nonumber
\end{eqnarray}
Here $\Phi _{{\bf mm}^{\prime }}^{{\bf nn}^{\prime }}(t)$ is a multiphonon
correlator, 
\begin{equation}
\Phi _{{\bf mm}^{\prime }}^{{\bf nn}^{\prime }}(t)\equiv \left\langle
\left\langle \hat{X}_{i}^{\dagger }(t)\hat{X}_{i^{\prime }}(t)\hat{X}%
_{j}^{\dagger }\hat{X}_{j^{\prime }}\right\rangle \right\rangle ,
\end{equation}
where 
\[
\hat{X}_{i}(t)=\prod_{_{{\bf q}}}\exp [u_{i}({\bf q,}t)d_{{\bf q}}-H.c.],
\]
and $u_{i}({\bf q,}t)=u_{i}({\bf q})e^{i\omega _{{\bf q}}t}$. $\ \hat{X}%
_{i}^{\dagger }(t)$ and $\hat{X}_{i^{\prime }}(t)$ commute for any $\gamma (%
{\bf q,}\nu )=\gamma (-{\bf q,}\nu )$ if ${\bf m\neq m}^{\prime }.$ Also $%
\hat{X}_{j}^{\dagger }$ and $\hat{X}_{j^{\prime }}$ commute, if ${\bf n\neq n%
}^{\prime }$, so that we can write 
\begin{eqnarray}
\hat{X}_{i}^{\dagger }(t)\hat{X}_{i^{\prime }}(t) &=&\prod_{{\bf q}%
}e^{[u_{i}({\bf q,}t)-u_{i^{\prime }}({\bf q,}t)]d_{{\bf q}}-H.c.]}, \\
\hat{X}_{j}^{\dagger }\hat{X}_{j^{\prime }} &=&\prod_{{\bf q}}e^{[u_{j}({\bf %
q})-u_{j^{\prime }}({\bf q})]d_{{\bf q}}-H.c.]}.
\end{eqnarray}
Applying twice the identity 
\[
e^{\hat{A}+\hat{B}}=e^{\hat{A}}e^{\hat{B}}e^{-[\hat{A},\hat{B}]/2},
\]
yields 
\begin{eqnarray}
\hat{X}_{i}^{\dagger }(t)\hat{X}_{i^{\prime }}(t)\hat{X}_{j}^{\dagger }\hat{X%
}_{j^{\prime }} &=&\prod_{{\bf q}}e^{\beta ^{\ast }d_{{\bf q}}^{\dagger
}}e^{-\beta d_{{\bf q}}}e^{-|\beta |^{2}/2}\times  \\
&&e^{[u_{i}({\bf q,}t)-u_{i^{\prime }}({\bf q,}t)][u_{j}^{\ast }({\bf q}%
)-u_{j^{\prime }}^{\ast }({\bf q})]/2-H.c.},  \nonumber
\end{eqnarray}
where 
\[
\beta =u_{i^{\prime }}({\bf q,}t)-u_{i}({\bf q,}t)+u_{j^{\prime }}({\bf q}%
)-u_{j}({\bf q}).
\]
Finally using the average \cite{fir} 
\[
\left\langle \left\langle e^{\beta ^{\ast }d_{{\bf q}}^{\dagger }}e^{-\beta
d_{{\bf q}}}\right\rangle \right\rangle =e^{-|\beta |^{2}n_{\omega }},
\]
where $n_{\omega }=[\exp (\omega _{{\bf q}}/T)-1]^{-1}$ is the Bose-Einstein
distribution function of phonons, we find 
\begin{eqnarray}
\Phi _{{\bf mm}^{\prime }}^{{\bf nn}^{\prime }}(t) &=&e^{-g^{2}({\bf m-m}%
^{\prime })}e^{-g^{2}({\bf n-n}^{\prime })}\times  \\
&&\exp \left\{ \frac{1}{2N}\sum_{{\bf q}}|\gamma ({\bf q})|^{2}F_{{\bf q}}(%
{\bf m,m}^{\prime },{\bf n,n}^{\prime })\frac{\cosh \left[ \omega _{{\bf q}%
}\left( \frac{1}{2T}-it\right) \right] }{\sinh \left[ \frac{\omega _{{\bf q}}%
}{2T}\right] }\right\} ,  \nonumber
\end{eqnarray}
where 
\begin{eqnarray}
F_{{\bf q}}({\bf m,m}^{\prime },{\bf n,n}^{\prime }) &=&\cos [{\bf q\cdot (n}%
^{\prime }-{\bf m})]+\cos [{\bf q\cdot (n}-{\bf m}^{\prime })]- \\
&&\cos [{\bf q\cdot (n}^{\prime }-{\bf m}^{\prime })]-\cos [{\bf q\cdot (n}-%
{\bf m})].  \nonumber
\end{eqnarray}
Taking into account that there are only bipolarons in the subspace, where $%
H_{b}$ operates, we finally rewrite the Hamiltonian in terms of the creation 
$b_{i}^{\dagger }=c_{{\bf m}\uparrow }^{\dagger }c_{{\bf m}\downarrow
}^{\dagger }$ and annihilation $b_{i}=c_{{\bf m}\downarrow }c_{{\bf m}%
\uparrow }$ singlet pair operators as 
\begin{eqnarray}
H_{b} &=&-\sum_{{\bf m}}\left[ \Delta +{\frac{1}{{2}}}\sum_{{\bf m^{\prime }}%
}v^{(2)}({\bf m-m}^{\prime })\right] n_{{\bf m}}+ \\
&&\sum_{{\bf m\neq m^{\prime }}}\left[ t({\bf m-m}^{\prime })b_{{\bf m}%
}^{\dagger }b_{{\bf m^{\prime }}}+{\frac{1}{{2}}}\bar{v}({\bf m-m}^{\prime
})n_{{\bf m}}n_{{\bf m^{\prime }}}\right] .  \nonumber
\end{eqnarray}
There are no triplet pairs in the Holstein model, because the Pauli
exclusion principle does not allow two electrons with the same spin occupy
the same site. Here $n_{{\bf m}}=b_{{\bf m}}^{\dagger }b_{{\bf m}}$ is the
bipolaron site-occupation operator, 
\begin{equation}
\bar{v}({\bf m-m}^{\prime })=4v({\bf m-m}^{\prime })+v^{(2)}({\bf m-m}%
^{\prime }),
\end{equation}
is the bipolaron-bipolaron interaction including the direct polaron-polaron
interaction $v({\bf m-m}^{\prime })$ and a repulsive correction of the
second order in $T({\bf m}),$ 
\begin{equation}
v^{(2)}({\bf m-m}^{\prime })=2i\int_{0}^{\infty }dte^{-i\Delta t}\Phi _{{\bf %
mm^{\prime }}}^{{\bf m^{\prime }m}}(t).
\end{equation}
This additional repulsion appears because a virtual hop of one of two
polarons of the pair is forbidden, if the neighbouring site is occupied by
another pair. The bipolaron transfer integral is of the second order in $T(%
{\bf m})$ 
\begin{equation}
t({\bf m-m}^{\prime })=-2iT^{2}({\bf m-m}^{\prime })\int_{0}^{\infty
}dte^{-i\Delta t}\Phi _{{\bf mm^{\prime }}}^{{\bf mm^{\prime }}}(t).
\end{equation}
The {\it bipolaronic} Hamiltonian, Eq.(31) describes the low-energy physics
of strongly coupled electrons and phonons. We use the explicit form of the
multiphonon correlator, Eq.(29), to calculate $t({\bf m})$ and $v^{(2)}({\bf %
m})$. If the phonon frequency is dispersionless, $\omega _{{\bf q}}=\omega
_{0},$ we obtain 
\begin{eqnarray*}
\Phi _{{\bf mm^{\prime }}}^{{\bf mm}^{\prime }}(t) &=&e^{-2g^{2}({\bf %
m-m^{\prime })}}\exp \left[ -2g^{2}({\bf m-m^{\prime })}e^{-i\omega _{0}t}%
\right] , \\
\Phi _{{\bf mm^{\prime }}}^{{\bf m}^{\prime }{\bf m}}(t) &=&e^{-2g^{2}({\bf %
m-m^{\prime })}}\exp \left[ 2g^{2}({\bf m-m^{\prime })}e^{-i\omega _{0}t}%
\right] 
\end{eqnarray*}
at $T\ll \omega _{0}.$ Expanding the time dependent exponents in the Fourier
series and calculating the integrals in Eqs.(34) and (33) yield \cite
{alekab0} 
\begin{equation}
t({\bf m)}=-{\frac{2T^{2}({\bf m})}{{\Delta }}}e^{-2g^{2}({\bf m)}%
}\sum_{l=0}^{\infty }\frac{[-2g^{2}({\bf m)}]^{l}}{l{!(1+l\omega _{0}/\Delta
)}}
\end{equation}
and 
\begin{equation}
v^{(2)}({\bf m})={\frac{2T^{2}({\bf m})}{{\Delta }}}e^{-2g^{2}({\bf m)}%
}\sum_{l=0}^{\infty }\frac{[2g^{2}({\bf m)}]^{l}}{l{!(1+l\omega _{0}/\Delta )%
}}.
\end{equation}
When $\Delta \ll \omega _{0},$ we can keep the first term only with $l=0$ in
the bipolaron hopping integral, Eq.(35). In this case the bipolaron
half-bandwidth $zt({\bf a)}$ is of the order of $2w^{2}/(z\Delta )$.
However, if the bipolaron binding energy is large, $\Delta \gg \omega _{0},$
the bipolaron bandwidth dramatically decreases proportional to $%
e^{-4g^{2}}\lll 1$ in the limit $\Delta \rightarrow \infty $. However, this
limit is not realistic because $\Delta =2E_{p}-V_{c}<2g^{2}\omega _{0}$. In
a more realistic regime, $\omega _{0}<\Delta <2g^{2}\omega _{0}$, Eq.(35)
yields 
\begin{equation}
t({\bf m)}\approx {\frac{2\sqrt{2\pi }T^{2}({\bf m})}{\sqrt{\omega
_{0}\Delta }}}\exp \left[ -2g^{2}-{\frac{\Delta }{{\omega }_{0}}}\left(
1+\ln \frac{{2g^{2}({\bf m)}\omega _{0}}}{\Delta }\right) \right] .
\end{equation}
On the contrary, the bipolaron-bipolaron repulsion, Eq.(36) has no small
exponent in the limit $\Delta \rightarrow \infty $, $v^{(2)}\varpropto
D^{2}/\Delta .$ Together with the direct Coulomb repulsion the second order $%
v^{(2)}$ ensures stability of the bipolaronic liquid against clustering.

\subsection{Intersite bipolaron band in a chain model}

Onsite bipolarons are very heavy for realistic values of the onsite
attractive energy $2E_{p}$ and phonon frequencies. Indeed, to bind two
polarons on a single site, $2E_{p}$ should overcome the onsite Coulomb
energy, which is typically of the order of $1$ $eV$ or higher. Optical
phonon frequencies are about $0.1\div 0.2$ $eV$ in novel superconductors
like oxides and doped fullerenes. Therefore in the framework of the Holstein
model, the mass enhancement exponent of onsite bipolarons in Eq.(37), is
rather large ($\gtrsim $ $\exp (2E_{p}/\omega _{0})>150),$ so that onsite
bipolarons could hardly account for high values of the superconducting
critical temperature \cite{ale2}.

But the Holstein model is not a typical model. The Fr\"{o}hlich interaction
with optical phonons, which is unscreened in polaronic systems, is much
stronger. This longer-range interaction leads to a lighter polaron in the
strong-coupling regime. Indeed, the polaron is heavy because it has to carry
the lattice deformation with it, the same deformation that forms the polaron
itself. Therefore, there exists a generic relation between the polaron
stabilization energy, $E_{p}$, and the renormalization of its mass, $%
m\propto \exp {(\gamma E_{p}/\omega }_{0}{),}$ where the numeric coefficient 
$\gamma $ depends on the radius of the interaction. For a short-range e-ph
interaction, the {\em entire} lattice deformation disappears and then forms
at another site, when the polaron moves between the nearest lattices sites.
Therefore, $\gamma =1$ and polarons and on-site bipolarons are very heavy
for the characteristic values of $E_{p}$ and ${\omega }_{0}$. On the
contrary, in case of a long-range interaction, only a fraction of the total
deformation changes every time the polaron moves and $\gamma $ could be as
small as $0.25$ \cite{ale2}. Clearly, this results in a dramatic
enlightening of the polaron since $\gamma $ enters the exponent. Thus the
small polaron mass could be $\leq 10\,m_{e}$ where a Holstein-like estimate
would yield a huge mass $10,000\,m_{e}$. The lower mass has important
consequences, because lighter polarons are more likely to remain mobile and
less likely to trap on impurities.

The bipolaron also becomes much lighter, if the e-ph interaction is
longer-range. There are two reasons for lowering of its mass with increasing
radius of the e-ph interaction. The first one is the same as in the case of
a single polaron discussed above. The second reason is the possibility to
form {\it intersite }bipolarons, which, in certain lattice structures,
tunnell coherently already in the first-order in $T({\bf m})$ \cite{ale2} ,
in contrast with onsite bipolarons, which tunnel only in the second-order,
Eq.(35).

To illustrate the essential dynamic properties of bipolarons formed by the
longer-range e-ph interaction let us discuss a few simplified models.
Following Bon\v{c}a and Trugman\cite{bon} we first consider a single
bipolaron in the chain model of Ref.\cite{alekor}. One can further simplify
the chain model by placing ions in the interstitial sites located between
Wannier orbitals of one chain, and allowing for the e-ph interaction only
with the nearest neighbours of another chain, as shown in Fig.1. The Coulomb
interaction is represented by the on-site Hubbard $U$ term. The model
Hamiltonian is 
\begin{eqnarray}
H &=&T(a)\sum_{j}[c_{j+1,s}^{\dagger }c_{js}+H.c.]+\omega
_{0}\sum_{i,j,s}g(i,j)\hat{n}_{js}(d_{i}^{\dagger }+d_{i})+ \\
&&\omega _{0}\sum_{i}[d_{i}^{\dagger }d_{i}+1/2]+U\sum_{i}\hat{n}_{j\uparrow
}\hat{n}_{j\downarrow }  \nonumber
\end{eqnarray}
in the site representation for both electrons and phonons , where 
\[
g(i,j)=g_{0}[\delta _{i,j}+\delta _{i,j+1}],
\]
and $i$, $j$ are integers sorting the ions and the Wannier sites,
respectively. This model is referred as the {\it extended }Holstein-Hubbard
model (EHHM) \cite{bon}. We can view the EHHM as the simplest model with
longer range than Holstein interaction. In comparison to the Fr\"{o}hlich
interaction the EHHM lacks long-range tail in the e-ph interaction, but
reveals the similar physical properties. In the momentum representation the
model is a one-dimensional case of the generic Hamiltonian, Eq.(1), with 
\begin{equation}
\gamma (q)=g_{0}\sqrt{2}(1+e^{iqa})
\end{equation}
and $\omega (q)=\omega _{0}.$ Using Eqs.(10), (14) and (8) we obtain 
\begin{equation}
E_{p}=\frac{g_{0}^{2}\omega _{0}a}{\pi }\int_{-\pi /a}^{\pi /a}dq[1+\cos
qa]=2g_{0}^{2}\omega _{0}
\end{equation}
for the polaron level shift, 
\begin{equation}
g^{2}=\frac{g_{0}^{2}a}{\pi }\int_{-\pi /a}^{\pi /a}dq[1-\cos
^{2}qa]=g_{0}^{2}
\end{equation}
for the mass enhancement exponent, and 
\begin{eqnarray}
v(0) &=&U-4g_{0}^{2}\omega _{0}, \\
v(a) &=&-\frac{2g_{0}^{2}\omega _{0}a}{\pi }\int_{-\pi /a}^{\pi /a}dq[1+\cos
qa]\cos qa=-2g_{0}^{2}\omega _{0}  \nonumber
\end{eqnarray}
for the onsite and intersite polaron-polaron interactions, respectively.
Hence the EHHM has the numerical coefficient $\gamma =1/2$, and the polaron
mass 
\begin{equation}
m_{EHP}^{\ast }\varpropto \exp \left( \frac{E_{p}}{2\omega _{0}}\right) 
\end{equation}
scales as the square root of the small Holstein polaron mass, $m_{SHP}^{\ast
}\varpropto \exp (E_{p}/\omega _{0}).$ In the case when $U<2g_{0}^{2}\omega
_{0}$, the onsite bipolaron has the lowest energy because $|v(0)|>|v(a)|.$
In this regime the bipolaron binding energy is 
\begin{equation}
\Delta =4g_{0}^{2}\omega _{0}-U.
\end{equation}
Using expression (34) for the bipolaron hopping integral we obtain the
bipolaron mass as 
\begin{equation}
m_{EHB}^{\ast \ast }\varpropto \exp \left( \frac{2E_{p}}{\omega _{0}}\right)
,
\end{equation}
if $\Delta \gg \omega _{0}.$ It scales as $(m_{EHP}^{\ast }/m)^{4}$ , but
occurs to be much smaller than the onsite bipolaron mass in the Holstein
model, $m_{SHB}^{\ast \ast }\varpropto \exp (4E_{p}/\omega _{0})$, which
scales as $(m_{SHP}^{\ast }/m)^{4}.$ In the opposite regime, when $%
U>2g_{0}^{2}\omega _{0},$ the intersite bipolaron has the lowest energy. Its
binding energy 
\begin{equation}
\Delta =2g_{0}^{2}\omega _{0}
\end{equation}
does not depend on $U.$ Different from the onsite singlet bipolaron, the
intersite bipolaron has four spin states, one singlet $S=0$ and three
triplet states, $S=1,$ with different $z$-components of the total spin, $%
S_{z}=0,\pm 1.$ In the chain model, Fig.1, the intersite bipolaron also
tunnells only in the second order in $T(a),$ when one of the electrons
within the pair hops to the left (right) and then the other follows. This
tunnelling involves the multiphonon correlation function $\Phi
_{j+1,j}^{j+2,j+1}$, Eq.(29), 
\[
\Phi _{j+1,j}^{j+2,j+1}=e^{-2g_{0}^{2}}.
\]
Hence the intersite bipolaron mass enhancement is 
\begin{equation}
m_{EHB}^{\ast \ast }\varpropto T^{-2}(a)\exp \left( \frac{E_{p}}{\omega _{0}}%
\right) \varpropto \left( \frac{m_{EHP}^{\ast }}{m}\right) ^{2}
\end{equation}
in the infinite Hubbard $U$ limit, $U\rightarrow \infty .$ We see that the
intersite bipolaron in the chain model is lighter than the onsite bipolaron,
but still remains much heavier than the polaron.

\bigskip

\subsection{Superlight intersite bipolarons in a ladder}

Any realistic theory of doped ionic insulators must include both the
long-range Coulomb repulsion between carriers and the strong long-range
electron-phonon interaction. From theoretical standpoint, the inclusion of
the long-range Coulomb repulsion is critical in ensuring that the carriers
would not form clusters. Indeed, in order to form stable bipolarons, the
el-ph interaction has to be strong enough to overcome the Coulomb repulsion
at short distances. Since the el-ph interaction is long-range, there is a
potential possibility for clustering. The inclusion of the Coulomb repulsion 
$V_{c}$ makes the clusters unstable. More precisely, there is a certain
window of $V_{c}/E_{p}$ inside which the clusters are unstable but
bipolarons nonetheless form. In this parameter window bipolarons repel each
other and propagate in a narrow band. At a weaker Coulomb interaction the
system is a charge segregated insulator, and at a stronger Coulomb repulsion
the system is the Fermi liquid, or the Luttinger liquid, if it is
one-dimensional.

Let us now apply a generic ``Fr\"{o}hlich-Coulomb'' Hamiltonian, which
explicitly includes the infinite-range Coulomb and electron-phonon
interactions, to a particular lattice structure \cite{alekor2}. The
implicitly present infinite Hubbard $U$ prohibits double occupancy and
removes the need to distinguish the fermionic spin. Introducing spinless
fermion operators $c_{{\bf n}}$ and phonon operators $d_{{\bf m}\nu }$, the
Hamiltonian is written as 
\begin{eqnarray}
H &=&\sum_{{\bf n\neq n^{\prime }}}T({\bf n-n^{\prime }})c_{{\bf n}%
}^{\dagger }c_{{\bf n^{\prime }}}+\sum_{{\bf n\neq n^{\prime }}}V_{c}({\bf %
n-n^{\prime }})c_{{\bf n}}^{\dagger }c_{{\bf n}}c_{{\bf n^{\prime }}%
}^{\dagger }c_{{\bf n^{\prime }}}+ \\
&&\omega _{0}\sum_{{\bf n\neq m,}\nu }g_{\nu }({\bf m-n})({\bf e}_{\nu
}\cdot {\bf e}_{{\bf m-n}})c_{{\bf n}}^{\dagger }c_{{\bf n}}(d_{{\bf m}\nu
}^{\dagger }+d_{{\bf m}\nu })+  \nonumber \\
&&\omega _{0}\sum_{{\bf m},\nu }\left( d_{{\bf m}\nu }^{\dagger }d_{{\bf m}%
\nu }+\frac{1}{2}\right) .  \nonumber
\end{eqnarray}
Here the e-ph and phonon terms are written in real space, which is more
convenient in working with complex lattices, $g_{\nu }({\bf m-n})$ is a
dimensionless {\em force} acting between the electron on site{\bf \ }${\bf n}
$ and the displacement of ion ${\bf m}$, ${\bf e}_{{\bf m-n}}\equiv ({\bf m-n%
})/|{\bf m-n}|$ is the unit vector in the direction from the electron ${\bf m%
}$ to the ion ${\bf n,}$ and ${\bf e}_{\nu }$ is the polarization vector of
the phonon branch $\nu $. Atomic orbitals of an ion adiabatically follow its
motion. Therefore the electron does not interact with the displacement of
the ion, whose orbital it occupies, that is $g_{\nu }(0)=0$.

In general, the many-body model Eq.(48) is of considerable complexity.
However, we are interested in the limit of the strong el-ph interaction. In
this case, the kinetic energy is a perturbation and the model can be grossly
simplified using the canonical transformation of Section 2, which has the
following form in the Wannier representation for electrons and phonons, 
\[
S=\sum_{{\bf m\neq n,}\nu }g_{\nu }({\bf m-n})({\bf e}_{\nu }\cdot {\bf e}_{%
{\bf m-n}})c_{{\bf n}}^{\dagger }c_{{\bf n}}(d_{{\bf m}\nu }^{\dagger }-d_{%
{\bf m}\nu }).
\]
The transformed Hamiltonian is 
\begin{eqnarray}
\tilde{H} &=&e^{S}He^{-S}=\sum_{{\bf n\neq n^{\prime }}}\hat{\sigma}_{{\bf %
nn^{\prime }}}c_{{\bf n}}^{\dagger }c_{{\bf n^{\prime }}}+\omega _{0}\sum_{%
{\bf m}\alpha }\left( d_{{\bf m}\nu }^{\dagger }d_{{\bf m}\nu }+\frac{1}{2}%
\right) + \\
&&\sum_{{\bf n\neq n^{\prime }}}v({\bf n-n^{\prime }})c_{{\bf n}}^{\dagger
}c_{{\bf n}}c_{{\bf n^{\prime }}}^{\dagger }c_{{\bf n^{\prime }}}-E_{p}\sum_{%
{\bf n}}c_{{\bf n}}^{\dagger }c_{{\bf n}}.  \nonumber
\end{eqnarray}
The last term describes the energy which polarons gain due to el-ph
interaction. $E_{p}$ is the familiar polaron level shift 
\begin{equation}
E_{p}=\omega \sum_{{\bf m}\nu }g_{\nu }^{2}({\bf m-n})({\bf e}_{\nu }\cdot 
{\bf e}_{{\bf m-n}})^{2},
\end{equation}
which is independent of ${\bf n}$. The third term on the right-hand side in
Eq.(49) is the polaron-polaron interaction: 
\begin{equation}
v({\bf n-n^{\prime }})=V_{c}({\bf n-n^{\prime }})-V_{ph}({\bf n-n^{\prime }}%
),
\end{equation}
where 
\begin{eqnarray*}
V_{ph}({\bf n-n^{\prime }}) &=&2\omega _{0}\sum_{{\bf m,}\nu }g_{\nu }({\bf %
m-n})g_{\nu }({\bf m-n^{\prime }})\times  \\
&&({\bf e}_{\nu }\cdot {\bf e}_{{\bf m-n}})({\bf e}_{\nu }\cdot {\bf e}_{%
{\bf m-n^{\prime }}}).
\end{eqnarray*}
The phonon-induced interaction $V_{ph}$ is due to displacements of common
ions by two electrons. Finally, the transformed hopping operator $\hat{\sigma%
}_{{\bf nn^{\prime }}}$ in the first term in Eq.(49) is given by 
\begin{eqnarray}
\hat{\sigma}_{{\bf nn^{\prime }}} &=&T({\bf n-n^{\prime }})\exp \left[ \sum_{%
{\bf m,}\nu }\left[ g_{\nu }({\bf m-n})({\bf e}_{\nu }\cdot {\bf e}_{{\bf m-n%
}})\right. \right.  \\
&&-\left. \left. g_{\nu }({\bf m-n^{\prime }})({\bf e}_{\nu }\cdot {\bf e}_{%
{\bf m-n^{\prime }}})\right] (d_{{\bf m}\nu }^{\dagger }-d_{{\bf m}\nu })%
\right] .  \nonumber
\end{eqnarray}
This term is a perturbation at large $\lambda $. Here we consider a
particular lattice structure (ladder), where bipolarons tunnell already in
the first order in $T({\bf n})$, so that $\hat{\sigma}_{{\bf nn^{\prime }}}$
can be averaged over phonons. When $T\lesssim \omega _{0}$ the result is 
\begin{equation}
t({\bf n-n^{\prime }})\equiv \left\langle \left\langle \hat{\sigma}_{{\bf %
nn^{\prime }}}\right\rangle \right\rangle _{ph}=T({\bf n-n^{\prime }})\exp
[-g^{2}({\bf n-n^{\prime }})],
\end{equation}
\begin{eqnarray*}
g^{2}({\bf n-n^{\prime }}) &=&\sum_{{\bf m},\nu }g_{\nu }({\bf m-n})({\bf e}%
_{\nu }\cdot {\bf e}_{{\bf m-n}})\times  \\
&&\left[ g_{\nu }({\bf m-n})({\bf e}_{\nu }\cdot {\bf e}_{{\bf m-n}})-g_{\nu
}({\bf m-n^{\prime }})({\bf e}_{\nu }\cdot {\bf e}_{{\bf m-n^{\prime }}})%
\right] .
\end{eqnarray*}
By comparing Eqs.(53) and Eq.(51), the mass renormalization exponent can be
expressed via $E_{p}$ and $V_{ph}$ as follows 
\begin{equation}
g^{2}({\bf n-n^{\prime }})=\frac{1}{\omega _{0}}\left[ E_{p}-\frac{1}{2}%
V_{ph}({\bf n-n^{\prime }})\right] .
\end{equation}
Now phonons are ``integrated out'' and the polaronic Hamiltonian is 
\begin{equation}
H_{p}=H_{0}+H_{pert},
\end{equation}
\[
H_{0}=-E_{p}\sum_{{\bf n}}c_{{\bf n}}^{\dagger }c_{{\bf n}}+\sum_{{\bf n\neq
n^{\prime }}}v({\bf n-n^{\prime }})c_{{\bf n}}^{\dagger }c_{{\bf n}}c_{{\bf %
n^{\prime }}}^{\dagger }c_{{\bf n^{\prime }}},
\]
\[
H_{pert}=\sum_{{\bf n\neq n^{\prime }}}t({\bf n-n^{\prime }})c_{{\bf n}%
}^{\dagger }c_{{\bf n^{\prime }}}.
\]
When $V_{ph}$ exceeds $V_{c}$ the full interaction becomes negative and
polarons form pairs.

The overall sign and magnitude of the interaction is given by the lattice
sum in Eq.(51) evaluation of which is elementary. Notice also, that
according to Eq.(54) an attractive interaction reduces the polaron mass (and
consequently bipolaron mass), while repulsive interaction enhances the mass.
Thus, the long-range character of the el-ph interaction serves the double
purpose. Firstly, it generates additional inter-polaron attraction. This
additional attraction helps overcome the direct Coulomb repulsion between
the polarons. Secondly, the Fr\"{o}hlich interaction makes the bipolarons
lighter.

The many-particle ground state of $H_{0}$ depends on the sign of the
polaron-polaron interaction, the carrier density, and the lattice geometry.
Here we consider the zig-zag ladder, Fig.2a, assuming that all sites are
isotropic two-dimensional harmonic oscillators. For simplicity, we also
adopt the nearest-neighbour approximation for both interactions, $g_{\nu }(%
{\bf l})\equiv g$, $V_{c}({\bf n})\equiv V_{c}$, and for the hopping
integrals, $T({\bf m})=T_{NN}$ for $l=n=m=a$, and zero otherwise. Hereafter
we set the lattice period $a=1$. There are four nearest neighbours in the
ladder, $z=4$. Then the {\it one-particle} polaronic Hamiltonian takes the
form 
\begin{eqnarray}
H_{p} &=&-E_{p}\sum_{n}(c_{n}^{\dagger }c_{n}+p_{n}^{\dagger }p_{n})+ \\
&&\sum_{n}[t^{\prime }(c_{n+1}^{\dagger }c_{n}+p_{n+1}^{\dagger
}p_{n})+t(p_{n}^{\dagger }c_{n}+p_{n-1}^{\dagger }c_{n})+H.c.],  \nonumber
\end{eqnarray}
where $c_{n}$ and $p_{n}$ are polaron annihilation operators on the lower
and upper sites of the ladder, respectively, Fig.2b. Using Eqs.(50) and (53)
we find 
\begin{eqnarray}
E_{p} &=&4g^{2}\omega _{0}, \\
t^{\prime } &=&T_{NN}\exp \left( -\frac{7E_{p}}{8\omega _{0}}\right) , 
\nonumber \\
t &=&T_{NN}\exp \left( -\frac{3E_{p}}{4\omega _{0}}\right) .  \nonumber
\end{eqnarray}
The Fourier transform of Eq.(56) yields 
\begin{eqnarray}
H_{p} &=&\sum_{k}(2t^{\prime }\cos k-E_{p})(c_{k}^{\dagger
}c_{k}+p_{k}^{\dagger }p_{k})+ \\
&&t\sum_{k}[(1+e^{ik})p_{k}^{\dagger }c_{k}+H.c.].  \nonumber
\end{eqnarray}
A linear transformation of $c_{k}$ and $p_{k}$ diagonalises the Hamiltonian,
so that the one-particle energy spectrum $E_{1}(k)$ is found from 
\begin{equation}
\det \left| \,\matrix{{2t'\cos k-E_{p}-E_{1}(k)}&{t(1+e^{ik})}\cr
{t(1+e^{-ik})}& {2t'\cos k-E_{p}-E_{1}(k)}\cr}\right| =0.
\end{equation}
There are two overlapping polaronic bands, 
\[
E_{1}(k)=-E_{p}+2t^{\prime }\cos (k)\pm 2t\cos (k/2) 
\]
with the effective mass $m^{\ast }=2/|4t^{\prime }\pm t|$ near their edges.

Let us now place two polarons on the ladder. The nearest neighbour
interaction, Eq.(51) is $v=V_{c}-E_{p}/2,$ if two polarons are on the
different sides of the ladder, and $v=V_{c}-E_{p}/4,$ if both polarons are
on the same side. The attractive interaction is provided via the
displacement of the lattice sites, which are the common nearest neighbours
to both polarons. There are two such nearest neighbours for the intersite
bipolaron of the type $A$ or $B$, Fig.2c, but there is only one common
nearest neighbour for the bipolaron $C$, Fig.2d. When $V_{c}>E_{p}/2$, there
are no bound states and the multi-polaron system is a one-dimensional
Luttinger liquid. However, when $V_{c}<E_{p}/2$ and consequently $v<0$, the
two polarons are bound into an intersite bipolaron of the type $A$ or $B$.

It is quite remarkable that the bipolaron tunnelling in the ladder appears
already in the first order with respect to a single-electron tunnelling.
This case is different from both onsite bipolarons and from the intersite
chain bipolarons discussed above, where the bipolaron tunnelling was of the
second order in $T(a)$. Indeed, the lowest energy degenerate configurations $%
A$ and $B$ are degenerate. They are coupled by $H_{pert}.$ Neglecting all
higher-energy configurations, we can project the Hamiltonian onto the
subspace containing $A$, $B$, and empty sites. The result of such a
projection is a bipolaronic Hamiltonian 
\begin{equation}
H_{b}=\left( V_{c}-\frac{5}{2}E_{p}\right) \sum_{n}[A_{n}^{\dagger
}A_{n}+B_{n}^{\dagger }B_{n}]-t^{\prime }\sum_{n}[B_{n}^{\dagger
}A_{n}+B_{n-1}^{\dagger }A_{n}+H.c.],
\end{equation}
where $A_{n}=c_{n}p_{n}$ and $B_{n}=p_{n}c_{n+1}$ are intersite bipolaron
annihilation operators, and the bipolaron-bipolaron interaction is dropped
(see below). Its Fourier transform yields two {\it bipolaron} bands, 
\begin{equation}
E_{2}(k)=V_{c}-{\frac{5}{{2}}}E_{p}\pm 2t^{\prime }\cos (k/2).
\end{equation}
with a combined width $4|t^{\prime }|$. The bipolaron binding energy in zero
order with respect to $t,t^{\prime }$ is 
\begin{equation}
\Delta \equiv 2E_{1}(0)-E_{2}(0)=\frac{E_{p}}{2}-V_{c}.
\end{equation}
The bipolaron mass near the bottom of the lowest band, $m^{\ast \ast
}=2/t^{\prime }$, is 
\begin{equation}
m^{\ast \ast }=4m^{\ast }\left[ 1+0.25\exp \left( \frac{{E_{p}}}{8\omega _{0}%
}\right) \right] .
\end{equation}
The numerical coefficient $1/8$ ensures that $m^{\ast \ast }$ remains of the
order of $m^{\ast }$ even at large $E_{p}$. This fact combines with a weaker
renormalization of $m^{\ast }$, Eq.(57), providing a {\em superlight}
bipolaron.

In models with strong intersite attraction there is a possibility of
clasterization. Similar to the two-particle case above, the lowest energy of 
$n$ polarons placed on the nearest neighbours of the ladder is found as 
\begin{equation}
E_{n}=(2n-3)V_{c}-\frac{6n-1}{4}E_{p}
\end{equation}
for any $n\geq 3$. There are {\em no} resonating states for $n$-polaron
nearest neighbour configuration if $n\geq 3$. Therefore there is no
first-order kinetic energy contribution to their energy. $E_{n}$ should be
compared with the energy $E_{1}+(n-1)E_{2}/2$ of far separated $(n-1)/2$
bipolarons and a single polaron for odd $n\geq 3$, or with the energy of far
separated $n$ bipolarons for even $n\geq 4$. ``Odd'' clusters are stable if 
\begin{equation}
V_{c}<\frac{n}{6n-10}E_{p},
\end{equation}
and ``even'' clusters are stable if 
\begin{equation}
V_{c}<\frac{n-1}{6n-12}E_{p}.
\end{equation}
As a result we find that bipolarons repel each other and single polarons at $%
V_{c}>\frac{3}{8}E_{p}$. If $V_{c}$ is less than $\frac{3}{8}E_{p}$ then
immobile bound clusters of three and more polarons could form. One should
notice that at distances much larger than the lattice constant the
polaron-polaron interaction is always repulsive, and the formation of
infinite clusters, stripes or strings is impossible \cite{alekab2}.
Combining the condition of bipolaron formation and that of the instability
of larger clusters we obtain a window of parameters 
\begin{equation}
\frac{3}{8}E_{p}<V_{c}<\frac{1}{2}E_{p},
\end{equation}
where the ladder is a bipolaronic conductor. Outside the window the ladder
is either charge segregated into finite-size clusters (small $V_{c}$), or it
is a liquid of repulsive polarons (large $V_{c}$).

\section{Conclusions}

It was analytically establshed a long time ago \cite{tja,hol,fir} that small
polarons are itinerant quasiparticles existing in the Bloch states at
temperatures below the characteristic phonon frequency for any strength of
the electron-phonon coupling. Here we have reviewed more recent studies of
small bipolarons in ionic lattices which show that the long-range
Fr\"{o}hlich interaction leads to relatively light intersite small
bipolarons with the atomic size of the wave function, large binding energy
and a large size of the phonon cloud. They are Bloch waves with the
effective mass, which is smaller by a few orders of magnitude than the mass
of onsite bipolarons in the nondispersive Holstein model. As discussed in a
few original papers, reviews and books the bipolaron theory describes $T_{c}$
of many cuprates without any fitting parameters, their non Fermi-liquid
normal state and the non-BCS superconducting state, and predicts
single-particle spectral properties compatible with the tunnelling and ARPES
spectroscopies of the cuprates.

The author greatly appreciates stimulating discussions with A.R. Bishop,
A.M. Bratkovsky, J.T. Devreese, D.M. Eagles, Yu.A. Firsov, L.P. Gor'kov,
V.V. Kabanov, P.E. Kornilovitch, W.Y. Liang, K.A. M\"{u}ller, and S.A.
Trugman. This work has been supported by the Leverhulme Trust (grant
F/00261/H).

\centerline{\bf Figure captions}

Fig.1. {\bf \ }{\small Simplified chain model with two electrons on the
chain interacting with nearest-neighbour ions of another chain. Second-order
intersite bipolaron tunnelling is shown by arrows.}

Fig.2. {\small One-dimensional zig-zag ladder. (a) Initial ladder with the
bare hopping amplitude }$T(a)${\small . (b) Two types of polarons with their
respective deformations. (c) Two degenerate bipolaron configurations A and
B. (d) A different bipolaron configuration C, which energy is higher than
that of A and B.}

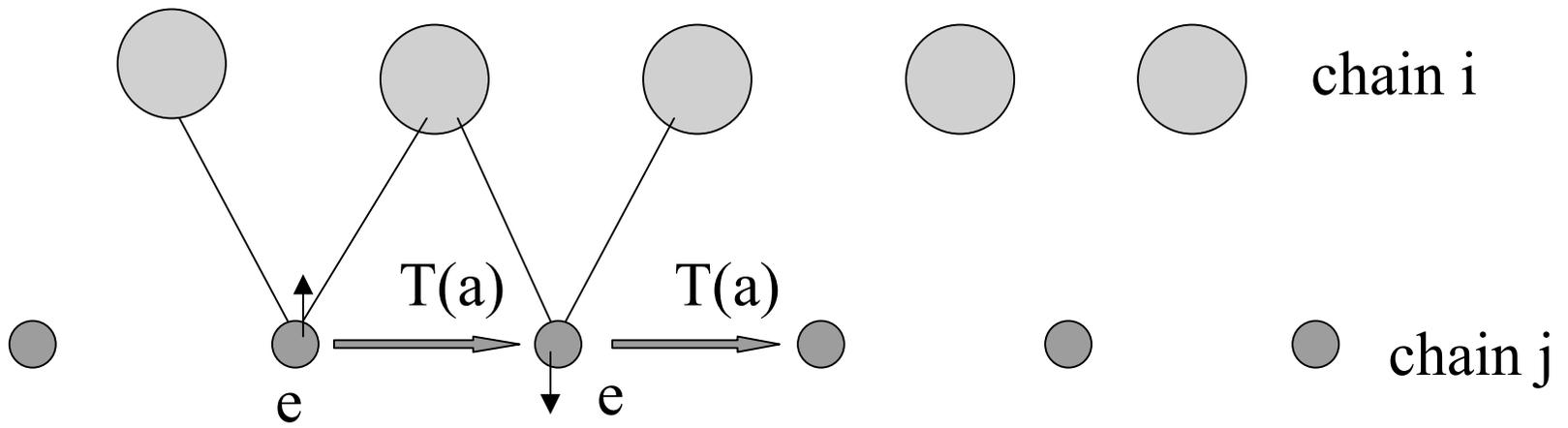
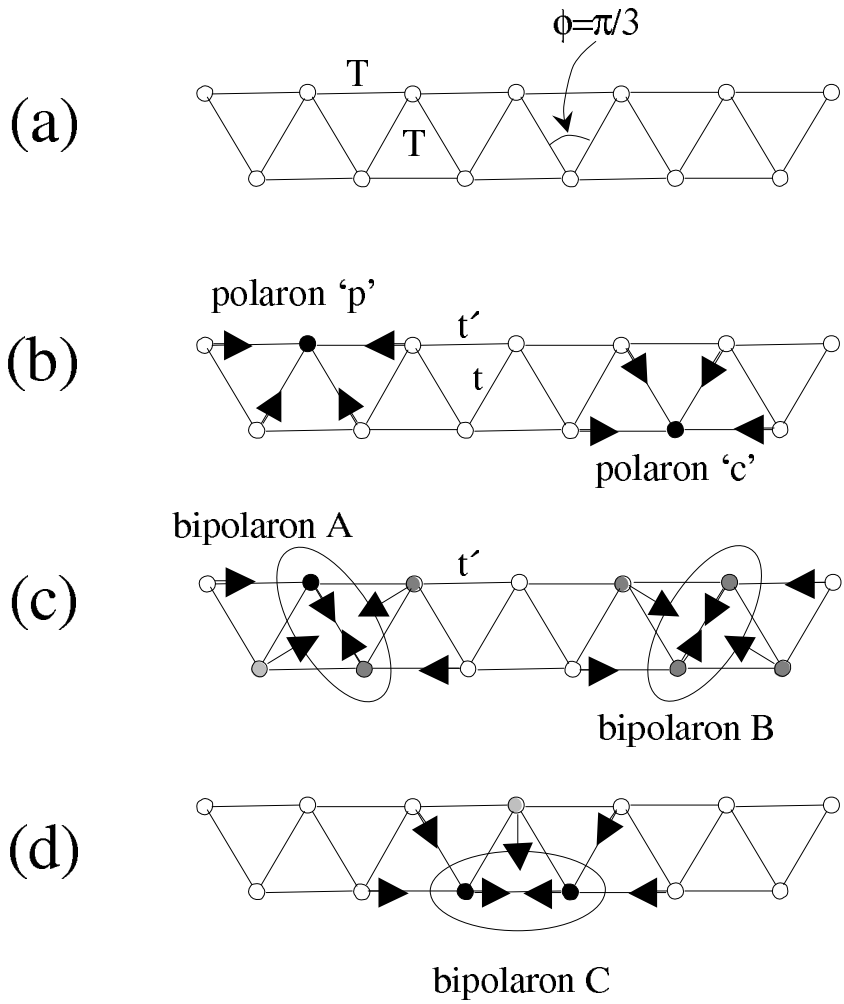
\end{document}